\documentclass[preprint,12pt]{elsarticle}

\usepackage[utf8]{inputenc}
\usepackage{amsmath,amssymb}
\usepackage{graphicx}
\usepackage{multicol}
\usepackage{hyperref}
\usepackage{url}
\usepackage{physics}
\usepackage{lipsum}
\usepackage{caption}
\usepackage{subcaption}
\usepackage{accents}
\usepackage{caption}
\usepackage{multirow}
\usepackage{float}
\usepackage{soul}
\usepackage{comment}
\usepackage{graphicx}
\usepackage{tabularx}
\usepackage{xcolor}

\newcommand{\m}{m_{\text{P}}}

\journal{Astroparticle Physics}

\begin{document}

\begin{frontmatter}



\title{Non-oscillating Early Dark Energy and Quintessence from $\alpha$-Attractors}


\author[inst1]{Lucy Brissenden}

\affiliation[inst1]{Consortium for Fundamental Physics, Physics Department,
Lancaster University, Lancaster LA1 4YB, United Kingdom.}

\author[inst1]{Konstantinos Dimopoulos}
\author[inst1]{Samuel S\'anchez L\'opez}
\ead{s.sanchezlopez@lancaster.ac.uk}

\begin{abstract}
Early dark energy (EDE) is one of the most promising possibilities in order to resolve the Hubble tension: the discrepancy between the locally measured and cosmologically inferred values of the Hubble constant. In this paper we propose a toy model of unified EDE and late dark energy (DE), driven by a scalar field in the context of $\alpha$-attractors. The field provides an injection of a subdominant dark energy component near matter-radiation equality, and redshifts away shortly after via free-fall, later refreezing to become late-time DE at the present day. Using reasonable estimates of the current constraints on EDE from the literature, we find that the parameter space is narrow but viable, making our model readily falsifiable. Since our model is non-oscillatory, the density of the field decays faster than the usual oscillatory EDE, thereby possibly achieving better agreement with observations.
\end{abstract}


\end{frontmatter}



\section{Introduction}
\label{sec:intro}

In the last few decades cosmological observations of the early and late Universe have converged into a broad understanding of the history of our Universe from the very first seconds of its existence until today. Thus, cosmology has developed a standard model called the concordance model, or in short $\rm\Lambda$CDM. 

However, the latest data might imply that the celebrated $\rm\Lambda$CDM model is not that robust after all. In particular, there is a 5$\sigma$ discrepancy between the estimated values of the current expansion rate, the Hubble constant $H_0$, as inferred by early Universe observations compared with late Universe measurements. This Hubble tension has undermined our confidence in $\rm\Lambda$CDM and as such it is investigated intensely at present.

In this work we study a toy model of unified early dark energy (EDE) and late dark energy (DE), which can simultaneously raise the inferred value of the Hubble constant $H_0$ coming from early-time data and explain the current accelerated expansion with no more tuning that in $\Lambda$CDM. We introduce a scalar field $\varphi$ in the context of $\alpha$-attractors, which is frozen at early times and unfreezes around matter-radiation equality, briefly behaving as a subdominant dark energy component to then undergo free-fall, redshifting away faster than radiation. At late times $\varphi$ behaves as quintessence. In contrast to most other works in the literature, ours is not an oscillating scalar field (see however Refs. \cite{Ye:2020oix,alphaattractorsede1,Lin:2019qug,Adil:2022hkj} for earlier attempts, the first two also in the context of $\alpha$-attractors).

We use natural units with $c=\hbar=1$, the reduced Planck mass $\m =
1/\sqrt{8\pi G} = 2.43 \times 10^{18}\text{GeV}$ and the $(-1,+1,+1,+1)$ sign convention for the metric throughout the present work.

\subsection{The Hubble tension}
Measurements in observational cosmology can broadly be classified into two groups. These are measurements of quantities which depend only on the early-time history of our Universe, such as the cosmic microwave background (CMB) radiation, emitted at redshift $z\simeq 1100$, or the baryon acoustic oscillations (BAO), and measurements of quantities which depend on present-day observations. A relevant example of the latter is the measurement of the distance to high-redshift type-Ia supernovae (SN Ia) by constructing a cosmic distance ladder \cite{sh0es2021}. This is achieved by starting with distances that can be directly resolved by using parallax and then moving to larger distances by using Cepheid variables and SN Ia. 

The value of the Hubble constant $H_0$ can in principle be inferred from early-time observations and directly obtained from late-time measurements. However, it has been found that while early-time results are in good agreement with each other, they disagree with current late-time data. Latest analysis of the CMB data gives the value inferred from the CMB temperature anisotropies power spectrum \cite{Planck2018} as
\begin{equation}
    H_{0} =  67.44\pm0.58 \ \textnormal{km} \ \textnormal{s}^{-1}\textnormal{Mpc}^{-1},
\end{equation}    
and a distance ladder measurement using Cepheid-SN Ia data 
from the SH0ES collaboration \cite{sh0es2021}
as 
\begin{equation}
H_{0} = 73.04\pm1.04 \ \textnormal{km} \ \textnormal{s}^{-1}\textnormal{Mpc}^{-1}.
\end{equation}
This is an $8\%$ difference between both values at a confidence level of $5\sigma$. It includes estimates of all systematic errors \cite{Verde:2019ivm} and the SH0ES team concludes that it has ``no indication of arising from measurement uncertainties or analysis variations considered to date''\footnote{Additionally, a closer study of SN-Ia results indicates the presence of a decreasing trend in $H_0$ with increasing redshift within datasets as well as between them \cite{Dainotti_2021,Dainotti_2022}, suggesting that the cause, whether systematic measurement error or theoretical, affects both datasets. Since there are likely to be fewer systematic errors that would affect both Planck and the cosmic distance ladder, this slightly increases the likelihood that theory holds the answer.}. It is becoming increasingly apparent with successive measurements that this tension is likely to have a theoretical resolution \cite{Dainotti_2021,Dainotti_2022,Mortsell:2018mfj,Escudero:2022rbq,Haridasu:2022dyp}, which can have many possible sources \cite{hubblehuntersguide,Gomez-Valent:2022bku} but increasingly favours early-time modifications \cite{Cai_2022latetime,Cai_2022matterperturb}.

\subsection{Early Dark Energy}
\label{subsec:ede}

One proposed class of solutions to the Hubble tension is models of EDE (the name was coined in Ref. \cite{Wetterich:2004pv} and early works include Refs. \cite{stringaxiverse,howearly,Calabrese_2011,Doran_2006}, 
followed by many others, \textit{e.g.} see Refs.~\cite{Sabla:2021nfy,Smith:2019ihp,Murai:2022zur,Capparelli:2019rtn,Berghaus:2022cwf,Berghaus:2019cls,Sakstein:2019fmf,Karwal:2021vpk,Sabla:2022xzj,Lin:2019qug,McDonough:2022pku,Poulin:2018dzj,newearlydarkenergy,doesnotrestoreconcordance,notexcluded,modellingeqnofstate,edecanresolve,chainede,hubblehuntersguide,rocknroll,alphaattractorsede1, Moshafi:2022mva,Guendelman:2022cop, TodaYo2021,Vagnozzi2020,Vagnozzi2022,Mohseni2022,abellán2023probing}). These involve an injection of energy in the dark energy sector at around the time of matter-radiation equality, which then is diluted
or otherwise decays away faster than the background energy density, such that it becomes negligible before it can be detected in current CMB observations. As briefly reviewed below, such models 
result in a slight change in the expansion history of the Universe, bumping up the value of the Hubble constant. 

It has previously been concluded \cite{Escudero:2022rbq,hubblehuntersguide, Gomez-Valent:2022bku} that EDE models 
are most likely to source a theoretical resolution to the Hubble tension.
One reason for this is that EDE 
can effect substantial modifications to $H_{0}$ without significant effect on other cosmological parameters, which are tightly constrained by observations\footnote{Models which modify other cosmological parameters are often unable to reconcile their changes with current observational constraints on said parameters (see Refs.~\cite{hubblehuntersguide}, \cite{Poulinreview2023} for a comprehensive review).}. 
In particular, 
EDE models can 
be incorporated into existing scalar-field models of inflation and late-time dark energy; one example of the latter 
is the model detailed in this work. 

However, precisely because EDE models exist so close in time to existing observational data, they are subject to significant constraints; the primary consideration being that EDE must be subdominant at all times and must decay away fast enough to be essentially negligible at the time of last scattering,  
translating to a redshift rate that is faster than radiation \cite{howearly}. So far, in previous works in EDE, this has been achieved via a variety of mechanisms, such as first or second-order phase transitions (\textit{e.g.} \cite{newearlydarkenergy}, \cite{rocknroll}). These 
abrupt events might have undesirable side-effects such as inhomogeneities from bubble collisions or topological defects. Other popular models \cite{alphaattractorsede1,hubblehuntersguide,stringaxiverse,howearly,newearlydarkenergy,doesnotrestoreconcordance,notexcluded,modellingeqnofstate,edecanresolve,chainede,rocknroll} typically feature oscillatory behaviour to achieve the rapid decay rate necessary for successful EDE. In this case, as with the original proposal in Ref. \cite{stringaxiverse}, the EDE field is taken to oscillate around its vacuum expectation value (VEV) in a potential minimum which is tuned to be of order higher than quartic. As a result, its energy density decays on average as $\propto a^{-n}$, with $4<n<6$. In contrast, in our model, the EDE scalar field experiences a period of kinetic domination, where the field is in non-oscillatory free-fall and its density decreases as $\rho \propto a^{-6}$, exactly rather than approximately. 

Before continuing, we briefly explain how EDE manages to increase the value of $H_0$ as from CMB observations.

Measurements of the CMB temperature anisotropies provide very tight constraints on a number of cosmological parameters. One would therefore think that this severely limits models which alter the Universe content and dynamics at this time. However, this is not the case for models that affect both the Hubble parameter and the comoving sound horizon $r_s$ (in this case during the drag epoch, shortly after recombination) while keeping the precisely-determined \cite{Planck2018} angle subtended by the sound horizon at last scattering $\theta_s$ unchanged. We remind the reader that the comoving sound horizon is given by

\begin{equation}
    r_s = \int^{\infty}_{z_{d}}\frac{c_{s}(z)}{H(z)}dz,
    \label{rs}
\end{equation}
where $c_{s}(z)$ is the sound speed of the baryon-photon fluid and $H(z)$ is the Hubble parameter, both as a function of redshift. An additional amount of dark energy in the Universe 
increases the total density, which in turn increases the Hubble parameter because of the Friedmann equation \mbox{$\rho\propto H^2$}. EDE briefly causes such an increase at or before matter-radiation equality, which lowers the value of the sound horizon because it increases $H(z)$ in Eq.~(\ref{rs}), leading to a larger value of $H_0$. This logic takes advantage of the fact that BAO measurements do not constrain the value of the sound horizon directly, but the combination $H(z)r_s$ \cite{BAOBOSS}. The same stands for CMB data, since the observations of the Planck satellite measure the quantity $\theta_{*} \equiv \frac{r_{s}(z_{*})}{D_{*}}$ \cite{Planck2013}, the angular scale of the sound horizon; given by ratio of the comoving sound horizon to the angular diameter distance at which we observe fluctuations. 
Both of these measurements entail an assumption of $\rm\Lambda$CDM cosmology and can be shown to be equally constrained by other models, provided that they make only small modifications which simultaneously lower the value of $r_s$ and increase $H_{0}$.

EDE may have a significant drawback in that it does not address the $\sigma_{8}$ tension (associated with matter clustering) and might in fact exacerbate it \cite{Escudero:2022rbq,deSa:2022hsh,Vagnozzi2021,Nunes2021}. 
However, recently several classes of models have emerged that alleviate both of these tensions simultaneously. These are axion models of coupled EDE and dark matter \cite{liu2023alleviating,liu2023kinetically,Lucca_2021,BeltranJimenez:2021wbq,patil2023coupled}. It is conceivable that an $\alpha$-attractor model such as ours could feature a similar interaction term.
\footnote{Moreover, tentative results indicate that previously neglected effects in galactic modelling may actually be responsible for the $\sigma_{8}$ tension.}

\subsection{$\alpha$-attractors}
\label{subsec:alphaattractors}

Our model unifies EDE with late DE in the context of $\alpha$-attractors\footnote{Earlier attempts for such unification in the same theoretical context can be found in Refs.~\cite{Ye:2020oix,Adil:2022hkj}.}. $\alpha$-attractors \cite{Kallosh:2013hoa, Linde:2018hmx, Kallosh:2015lwa, Cecotti:2014ipa, lindealphaattractors,Ferrara:2013rsa, Ferrara:2013eqa, Ferrara:2014rya, Kallosh:2014rga} appear naturally in conformal field theory or supergravity theories. They are a class of models whose inflationary predictions continuously interpolate between those of chaotic inflation \cite{Linde:2014hfa} and those of Starobinsky \cite{Starobinsky:1980te} and Higgs inflation \cite{Bezrukov:2007ep}. In supergravity, introducing curvature to the internal field-space manifold can give rise to a non-trivial K\"{a}hler metric, which results in kinetic poles for some of the scalar fields of the theory. The free parameter $\alpha$ is inversely proportional to said curvature. It is also worth clarifying what is meant by the word ``attractor''. It is
used to refer to the fact that the inflationary predictions are largely insensitive of the specific characteristics of the potential under consideration. Such an attractor behaviour is seen for sufficiently large curvature (small $\alpha$) in the internal field-space manifold. 

In practical terms, the scalar field has a non-canonical kinetic term, featuring two poles, which the field cannot transverse. To aid our intuition, the field can be canonically normalised via a field redefinition, such that the finite poles for the non-canonical field are transposed to infinity for the canonical one. As a result, the scalar potential is ``stretched'' near the poles, resulting in two plateau regions, which are useful for modelling inflation, see \textit{e.g.} Refs. \cite{Alho:2017opd, Odintsov:2016vzz, Braglia:2022phb, Kallosh:2022ggf, Achucarro:2017ing, Iarygina:2020dwe}, or quintessence \cite{Akrami:2017cir}, or both, in the context of quintessential inflation \cite{Akrami:2017cir,Dimopoulos:2017tud, Dimopoulos:2017zvq}.

Following the standard recipe, we introduce two poles at $\varphi=\pm\sqrt{6\alpha}\,m_P$ by considering the Lagrangian 

\begin{equation}
    \mathcal{L} = \frac{-\frac{1}{2}(\partial\varphi)^{2}}{(1-\frac{\varphi^{2}}{6\alpha \,\m^{2}})^{2}} - U(\varphi)\,,
    \label{L0}
\end{equation}
where $\varphi$ is the non-canonical scalar field and we use the short-hand notation $(\partial \varphi)^2\equiv g^{\mu\nu}\partial_{\mu}\varphi\,\partial_{\nu}\varphi$.
We then redefine the non-canonical field in terms of the canonical scalar field $\phi$ as
\begin{equation} 
    \text{d}\phi =\frac{\text{d}\varphi}{1-\frac{\varphi^2}{6\alpha \m^2}} \Rightarrow\quad\varphi = \m\sqrt{6\alpha}\,\tanh{\left(\frac{\phi}{\sqrt{6\alpha}\,\m}\right)}\,.
    \label{phivarphi}
\end{equation}
It is obvious that the poles \mbox{$\varphi=\pm \sqrt{6\alpha}\,\m$} are transposed to infinity. 

In terms of the canonical field, the Lagrangian now reads

\begin{eqnarray}
    &&\mathcal{L} = -\frac{1}{2}(\partial\phi)^{2} - V(\phi),
    \label{L}
\end{eqnarray}
where
\mbox{$V(\phi)=U\left(\m\sqrt{6\alpha}\,\tanh{\left(\frac{\phi}{\sqrt{6\alpha}\,\m}\right)}\right).$}

\subsection{Quintessence}
\label{subsec:quintessence}

``Early'' dark energy is so named in order to make it distinct from ``late'' dark energy, which is the original source of the name (and often just called dark energy). In cosmological terms the latter is just beginning to dominate the Universe at present, making up approximately $70\%$ of the Universe's energy density
\cite{SupernovaSearchTeam:1998fmf}.
This is the mysterious unknown substance that is responsible for the current accelerating expansion of the Universe and has an equation-of-state (barotropic) parameter of 
$w=-1.03\pm0.03$ \cite{Planck2018}. 

Late DE that is due to an (as-yet-undiscovered) scalar field is called \textit{quintessence} \cite{Caldwell:1997ii}, so-named because it is the fifth element making up the content of the Universe \footnote{After baryonic matter, dark matter, photons and neutrinos.}. In this case, the Planck-satellite bound on the barotropic parameter of DE is
$-1\leq w<-0.95$ \cite{Planck2018}. Quintessence can be distinguished from a cosmological constant because a scalar field has a variable barotropic parameter and can therefore exhibit completely different behaviour in different periods of the Universe's history. In order to get it to look like late-time DE, a scalar field should be dominated by its potential density, making its barotropic parameter sufficiently close to $-1$. It is useful to consider the CPL parametrization, which is obtained by Taylor expanding $w(z)$ near the present as 
\cite{Chevallier:2000qy,Linder:2002et}
\begin{equation}
    w(z)=w_0+w_a\frac{z}{z+1}\,,
\end{equation}
where \mbox{$w_a\equiv-({\rm d}w/{\rm d}a)_0$}. The Planck satellite observations impose the bounds \cite{Planck2018}
\begin{eqnarray}
-1\leq w_{0}<-0.95 \quad \text{and} \quad w_a=-0.29^{+0.32}_{-0.26}.
  \label{CPL}
\end{eqnarray}

\section{The Model}
\label{sec:model}

\subsection{Lagrangian and Field Equations}
\label{subsec:lagrangian}

We consider a potential of the form 
\begin{eqnarray}
   U(\varphi) = V_X\exp(-\lambda e^{\kappa\varphi/\m}),
      \label{Vvarphi}
\end{eqnarray}
where
\begin{equation}
    V_\Lambda\equiv \exp(-\lambda e^{\kappa\sqrt{6\alpha}})V_X\,,
    \label{eqn:potedef}
\end{equation}
and $\alpha, \kappa, \lambda$ are dimensionless model parameters, $V_X$ is a constant energy density scale and $\varphi$ is the non-canonical scalar field with kinetic poles given by the typical alpha attractors form (see Ref. \cite{lindealphaattractors}) with a Lagrangian density given by Eq.~(\ref{L0}). In the above, $V_\Lambda$ is the vacuum density at present\footnote{The model parameter is $V_X$ and not $V_\Lambda$, the latter being related to $V_X$ and the remaining model parameters via Eq.~\eqref{eqn:potedef}.}.
To assist our intuition, we switch to the canonically normalised (canonical) scalar field $\phi$, using the transformation in Eq.~\eqref{phivarphi}. In terms of the canonical scalar field, the Lagrangian density is then given by Eq.~\eqref{L}, where the scalar potential is
\begin{equation}
    V(\phi) = \exp(\lambda e^{\kappa\sqrt{6\alpha}})V_{\Lambda}\exp[-\lambda e^{\kappa\sqrt{6\alpha}\tanh(\phi/\sqrt{6\alpha}\,\m)}]\,.
    \label{eqn:currentpotential}
\end{equation}

As usual, the Klein-Gordon equation of motion for the homogeneous canonical field is

\begin{equation}
    \ddot{\phi} + 3H\dot{\phi} + V'(\phi) = 0\,,
    \label{KG}
\end{equation}
where the dot and prime denote derivatives with respect to the cosmic time and the scalar field respectively, and we assumed that the field was homogenised by inflation, when the latter overcame
the horizon problem. 
\subsection{Shape of Potential and Expected Behaviour}
\label{subsec:expectedbehaviour}

Henceforth we will discuss the behaviour of the EDE scalar field in terms of the variation, \textit{i.e.} movement in field space, of the canonical field. 

\subsection{Asymptotic forms of the scalar potential}

We are interested in two limits for the potential in Eq.~\eqref{eqn:currentpotential}: \mbox{$\phi\rightarrow 0$} ($\varphi\rightarrow 0$) and
 \mbox{$\phi\rightarrow+\infty$} ($\varphi\rightarrow+\sqrt{6\alpha}\,m_P$).
 The first limit corresponds to matter-radiation equality. In this limit, the potential is

\begin{equation}
    V_{\textnormal{eq}} \simeq \exp[\lambda(e^{\kappa\sqrt{6\alpha}}-1)]  V_{\Lambda}\exp(-\kappa\lambda\,\phi_{\textnormal{eq}}/\m)\,,
    \label{eqn:lowphiapprox}
\end{equation}
where the subscript `eq' denotes
the time of matter-radiation equality when the field unfreezes. It is assumed that the field was originally frozen there. We discuss and justify this assumption in Sec.~\ref{sec:discussion}.

After unfreezing, it is considered that the field has not varied much, for the above approximation to hold, \textit{i.e.},

\begin{equation}
    0\lesssim\phi_{\textnormal{eq}}\ll \sqrt{6\alpha}\m\,.
    \label{eqn:condition}
\end{equation}
This is
a reasonable assumption given that the field begins shortly before matter-radiation equality frozen at the origin, unfreezing at some point during this time \footnote{There is no suggestion in the EDE literature \cite{alphaattractorsede1,hubblehuntersguide,stringaxiverse,howearly,newearlydarkenergy,doesnotrestoreconcordance,notexcluded,modellingeqnofstate,edecanresolve,chainede,rocknroll} that the field has to unfreeze at any particular time, as long as it does not grow to larger than the allowed fraction and its energy density is essentially negligible by the time of decoupling.}. 

At large $\phi$ ({\em \textit{i.e.}} $\phi \rightarrow \infty$), the non-canonical field is near the kinetic pole ($\varphi\rightarrow+\sqrt{6\alpha}\,\m$). Then the potential in this limit is

\begin{equation}
    V_{0} \simeq V_{\Lambda}\left[1+2\kappa\lambda e^{\kappa\sqrt{6\alpha}}\sqrt{6\alpha}\,\exp\left(-\frac{2\phi_{0}}{\sqrt{6\alpha}\,\m}\right)\right],
   \label{eqn:highphiapprox}
\end{equation}
which, even for sub-Planckian total field excursion in $\phi$, should be a good approximation for sufficiently small $\alpha$. The subscript `0' denotes the present time\footnote{Note that, as the field becomes sufficiently large, the potential approaches the positive constant $V_\Lambda$, which corresponds to non-zero vacuum density with $w=-1$, as in $\rm\Lambda$CDM. Thus, our model outperforms pure quintessence (with $-1<w<-0.95$ \cite{Planck2018}), which can push  $H_0$ to lower instead of higher values \cite{Banerjee:2020xcn,Lee:2022cyh}.}.

\begin{figure}[h]
    \centering
    \includegraphics[width=0.9\textwidth]{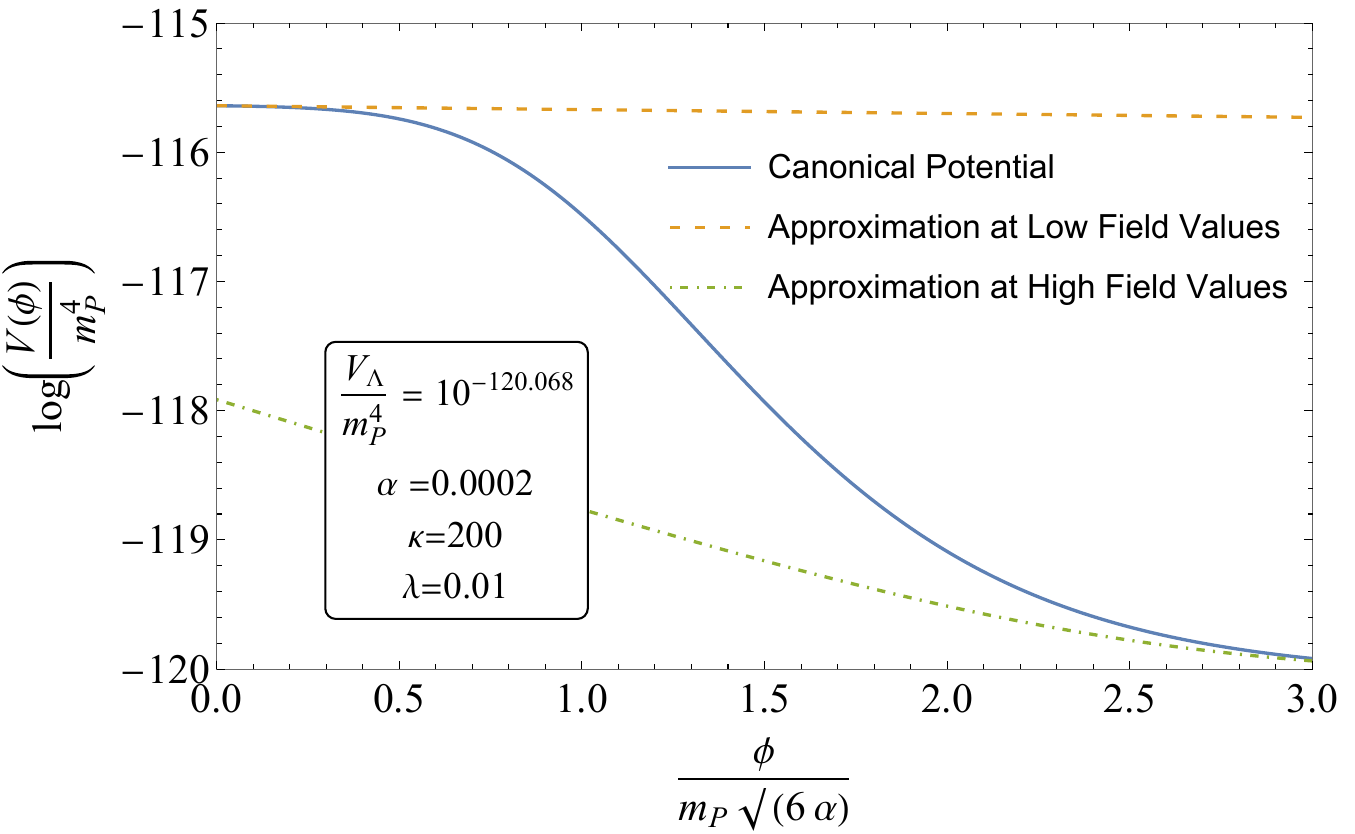}
    \caption{Graph of the canonical potential and its two approximations for small and large field values, given in Eqs.~\eqref{eqn:lowphiapprox}, \eqref{eqn:highphiapprox} respectively. These approximations are useful because they are simple exponential potentials with well-known attractors. It can be readily seen that, after leaving the origin, the field jumps off a potential plateau and is free-falling as a result.}
    \label{fig:canonicalapproximations}
\end{figure}

The above approximations describe well the scalar potential near equality and the present time, as shown in Fig.~\ref{fig:canonicalapproximations}. As we explain below, in between these regions, the scalar field free-falls and becomes oblivious of the scalar potential as the term $V'(\phi)$ in its equation of motion \eqref{KG} becomes negligible.

\subsubsection{Expected Field Behaviour}
\label{subsubsec:expectedbehaviour}

Here we explain the rationale behind the mechanism envisaged. We make a number of crude approximations, which enable us to follow the evolution of the scalar field, but which need to be carefully examined numerically. We do so in the next section.

First, we consider that originally the field is frozen at zero (for reasons explained in Section ~\ref{sec:discussion}).
Its energy density is such that
it remains frozen there until equality, when it thaws following the appropriate exponential attractor, since $V_{\rm eq}$ in Eq.~\eqref{eqn:lowphiapprox} is approximately exponential \cite{Copeland:1997et}. Assuming that this is the subdominant attractor requires that the strength of the exponential is \cite{Copeland:2006wr,kostasbook}
\begin{equation}
Z\equiv\kappa\lambda>\sqrt 3\,.
  \label{z}
  \end{equation}
The subdominant exponential attractor dictates that the energy density of the rolling scalar field mimics the dominant background energy density. Thus, the density parameter of the field is constant, given by the value \cite{Copeland:1997et,Copeland:2006wr,kostasbook}
\begin{equation}
  \mathit{\Omega}_\phi^{\rm eq}\simeq\frac{3}{Z^2}=\frac{3}{(\kappa\lambda)^2}<1
\label{Omegaphieq}
\end{equation}
This provides an estimate of the moment when the originally frozen scalar field, unfreezes and begins rolling down its potential. Unfreezing happens when $\mathit{\Omega}_\phi$ (which is growing while the field is frozen, because the background density decreases with the expansion of the Universe) obtains the above value.

However, after unfreezing, the field soon experiences the full exp(exp) steeper than exponential potential so, it does not follow the subdominant attractor any more but it
free-falls, \textit{i.e.}, its energy density is dominated by its kinetic component, such that its density scales as \mbox{$\rho_\phi\simeq\frac12\dot\phi^2\propto a^{-6}$}, until it refreezes at a larger value $\phi_F$. This value is estimated as follows.

In free-fall, the slope term in the equation of motion \eqref{KG} of the field is negligible, so that the equation is reduced to
\mbox{$\ddot\phi+3H\dot\phi\simeq0$}, where $H=2/3t$ after equality.
The solution is
\begin{equation}
\phi(t)=\phi_{\rm eq}+\frac{C}{t_{\rm eq}}\left(1-\frac{t_{\rm eq}}{t}\right)\,,
\label{phit}
\end{equation}
where $C$ is an integration constant. From the above, it is straightforward
to find that \mbox{$\dot\phi=Ct^{-2}$}. Thus, the density parameter at equality is
\begin{eqnarray}
    \mathit{\Omega}_\phi^{\rm eq} && =\left.\frac{\rho_\phi}{\rho}\right|_{\rm eq}=
  \frac{\frac12 C^2 t_{\rm eq}^{-4}}{\frac43(\frac{m_P}{t_{\rm eq}})^2}=\frac38\frac{C^2}{(m_Pt_{\rm eq})^2}\nonumber\\
&&\Rightarrow
C=\sqrt{\mbox{$\frac83$}\mathit{\Omega}_\phi^{\rm eq}}\,m_Pt_{\rm eq}=\frac{\sqrt 8}{\kappa\lambda}\,m_P \,t_{\rm eq}\;,
\label{C}
\end{eqnarray}
where we used Eq.~(\ref{Omegaphieq}), \mbox{$\rho_\phi\simeq\frac12\dot\phi^2$} and that
\mbox{$\rho=1/6\pi Gt^2=\frac43(m_P/t)^2$}. Thus, the field freezes at the value
\begin{equation}
  \phi_0=\phi_{\rm eq}+C/t_{\rm eq}=\phi_{\rm eq}+
  \frac{\sqrt 8}{\kappa\lambda}\,m_P\;,
\label{phi0}
\end{equation}
where we considered that \mbox{$t_{\rm eq}\ll t_{\rm freeze}<t_0$}\ .

Using that \mbox{$t_{\rm eq}\sim 10^4\,$y} and \mbox{$t_0\sim 10^{10}\,$y}, we can estimate
\begin{equation}
\begin{split}
  \frac{V_{\rm eq}}{V_0}&\simeq\frac{\Omega_\phi^{\rm eq}\rho_{\rm eq}}{0.7\,\rho_0}
  \simeq\frac{30}{7(\kappa\lambda)^2}
  \left(\frac{t_0}{t_{\rm eq}}\right)^2\\
  &\simeq
  \frac{3}{7(\kappa\lambda)^2}\times 10^{13}\,.
  \label{Vratio}
  \end{split}
\end{equation}
Now, from Eqs.~(\ref{eqn:lowphiapprox}), (\ref{eqn:highphiapprox}) we find
\begin{equation}
 \frac{V_{\rm eq}}{V_0}\simeq 
 \frac{e^{\lambda(e^{\kappa\sqrt{6\alpha}}-1)}\exp(-\kappa\lambda\,\phi_{\rm eq}/m_P)}{1+
   2\kappa\lambda\,e^{\kappa\sqrt{6\alpha}}\sqrt{6\alpha}
   \exp(-2\phi_0/\sqrt{6\alpha}\,m_P)}\,.
\end{equation}
In view of Eqs.~(\ref{eqn:condition}), (\ref{phi0}), the above can be written as
\begin{equation}
  \frac{V_{\rm eq}}{V_0}\simeq
  \frac{e^{\lambda(e^{\kappa\sqrt{6\alpha}}-1)}}{1+ 2\kappa\lambda\,
    e^{\kappa\sqrt{6\alpha}}\sqrt{6\alpha}\,e^{-2\sqrt 8/\kappa\lambda\sqrt{6\alpha}}}\,.
  \label{Vratio+}
\end{equation}

Taking \mbox{$\mathit{\Omega}_\phi^{\rm eq}\simeq 0.1$} as required by EDE, Eq.~(\ref{Omegaphieq}) suggests
\begin{equation}
  \kappa\lambda\simeq\sqrt{30}\,.
  \label{30}
\end{equation}
Combining this with Eq.~(\ref{Vratio}) we obtain
\begin{equation}
  e^{\frac{\sqrt{30}}{\kappa}(e^{\kappa\sqrt{6\alpha}}-1)}\sim 10^{12}/7\,,
\label{master}
\end{equation}  
where we have ignored the second term in the denominator of the right-hand-side of Eq.~(\ref{Vratio+}).

From the above we see that, $\kappa$ is large when $\alpha$ is small. Taking, as an example,
\mbox{$\alpha=0.01$} we obtain \mbox{$\kappa\simeq 18$} and
\mbox{$\lambda\simeq 0.30$} (from Eq.~(\ref{30})). With these values, the second term in the denominator of the right-hand-side of Eq.~(\ref{Vratio+}), which was ignored above, amounts to the
value 3.2. This forces a correction to the ratio $V_{\rm eq}/V_0$ of order unity, which means that the order-of-magnitude estimate in Eq.~\eqref{master} is not affected.

Using the selected values,
Eq.~(\ref{phi0}) suggests that the total excursion of the field is
\begin{equation}
  \Delta\phi=\phi_0-\phi_{\rm eq}=\frac{\sqrt 8}{\kappa\lambda}\,m_P
  \simeq 0.5\,m_P\;,
\end{equation}
\textit{i.e.} it is sub-Planckian. In the approximation of Eq.~(\ref{eqn:lowphiapprox}), we see that
the argument of the exponential becomes
\mbox{$\kappa\lambda\Delta\phi/m_P\simeq 2.7>1$}, where we used Eq.~(\ref{30}).
This means that the exponential
approximation breaks down and the exp(exp) potential is felt as considered, as depicted also in Fig.~\ref{fig:canonicalapproximations}.

For small $\alpha$, the eventual exponential potential in Eq.~\eqref{eqn:highphiapprox} is steep, which suggests
that field rushes towards the minimum at infinity. However, the barotropic parameter is
\mbox{$w\approx -1$} because the potential is dominated by the constant $V_\Lambda$.

\subsection{Tuning requirements}

Our model addresses in a single shot two cosmological problems: firstly, the Hubble tension between inferences of $H_{0}$ using early and late-time data; and secondly, the reason for the late-time accelerated expansion of the Universe; late DE. However, it is subject to some tuning. Namely, the two free parameters $\kappa$ and $\lambda$, the intrinsic field-space curvature dictated by $\alpha$, and the scale of the potential introduced by $V_{\Lambda}$. 

As we have seen $\kappa$ and $\lambda$ seem to take natural values, not too far from order unity. Regarding $\alpha$ we only need that it is small enough to lead to rapid decrease of the exponential contribution in the scalar potential in Eq.~\eqref{eqn:highphiapprox}, leaving the constant $V_\Lambda$ to dominate at present. We show in the next section that \mbox{$\alpha\sim 10^{-4}$} is sufficient for this task. This leaves $V_\Lambda$ itself.

The required tuning of this parameter is given by $V_{\Lambda} = 
\Big(\frac{H_{0}^{\rm Planck}}{H_{0}^{\rm SH0ES}}\Big)^{2}V_\Lambda^{\rm Planck}$, where \mbox{$V_\Lambda^{\rm Planck}=\Omega_\Lambda\rho_0$}. Since 
\mbox{$\Big(\frac{H_{0}^{\rm Planck}}{H_{0}^{\rm SH0ES}}\Big)^{2}\simeq(\frac{67.44}{73.04})^2=0.8525$} we see that the required fine-tuning of our $V_\Lambda$ is not different from the fine-tuning introduced in $\Lambda$CDM. However, in contrast to $\Lambda$CDM, our proposal addresses two cosmological problems; not only late DE but also the Hubble tension.

\section{Numerical Simulation}
\label{sec:numerics}

In order to numerically solve the background dynamics of the system, it is enough to solve for the scale factor $a(t)$, the field $\phi(t)$ and the background perfect fluid densities $\rho_{\text{m}}(t)$ and $\rho_{\text{r}}(t)$ (of matter and radiation respectively), as every other quantity depends on these. They are governed by the Friedmann equation, the Klein-Gordon equation and the continuity equations respectively. Of course, the Klein-Gordon equation is a second order ordinary differential equation, while the continuity equations are first order so that we need the initial value and velocity of $\phi$ and just the initial value of $\rho_{\text{m}}$ and $\rho_{\text{r}}$ as initial conditions. As described above, the field starts frozen and unfreezes around matter-radiation equality. Effectively, this means using $\phi_{\text{ini}}=0$ and $\Dot{\phi}_{\text{ini}}=0$ as initial conditions, while the initial radiation and matter energy densities are chosen to satisfy the bounds obtained by Planck \cite{Planck2018} at matter-radiation equality, \textit{i.e.}, scaled back from $\rho_{\text{m}}(t_{\text{eq}})=\rho_{\text{r}}(t_{\text{eq}})=1.27 \times 10^{-110}\m^4$, at some arbitrary redshift $z_{\rm ini}=10^4$.

For convenience,  we rewrite the equations in terms of the logarithmic energy densities $\Tilde{\rho}_{m}(t) = \ln{(\rho_{m}(t)/\m^4)}$ and $\Tilde{\rho}_{r}(t) = \ln{(\rho_{r}(t)/\m^4)}$. Plugging the first Friedmann equation in the Klein-Gordon equation, gives

\begin{equation}
\ddot{\phi}(t) + 
\frac{\sqrt{3\rho(t)}}{\m}\;\dot{\phi}(t) + \frac{dV}{d\phi} = 0,
\label{eqn:fieldeqn}
\end{equation}

\begin{equation}
    \dot{\Tilde{\rho}}_{m}(t) + \frac{\sqrt{3\rho(t)}}{\m} = 0,
    \label{eqn:mattereqnofstate}
\end{equation}

\begin{equation}
        \dot{\Tilde{\rho}}_{r}(t) + \frac43\frac{\sqrt{3\rho(t)}}{\m} = 0,
        \label{eqn:radiationeqnofstate}
\end{equation}
where \ $3\m^2H^2(t)=\rho(t) = 
\rho_{\phi}(t)+
[\,\exp(\Tilde{\rho}_{m}(t)) + \exp(\Tilde{\rho}_{r}(t))]\m^4 
$ \ and $\rho_{\phi}(t) = K(\phi(t)) + V(\phi(t))$ where $K(\phi(t)) = \frac{1}{2}\dot{\phi}^2(t)$ and $V(\phi(t))$ is given by Eq.~\eqref{eqn:currentpotential}.

As mentioned in Section \ref{sec:discussion}, we assume the field to be initially frozen at an ESP, such that it could have been the inflaton or a spectator field at earlier times. The time of unfreezing is then controlled only by the parameters of the potential.

The simulation is terminated when the density parameter of the field becomes equal to the density parameter of dark energy today $\Omega_{\Lambda}=0.6889$ \cite{Planck2018}.

 \break

\begin{table}[H]

\centering

\begin{tabular}{|p{3.0cm}|p{1.6cm}|p{3.0cm}|p{4.0cm}|}
\hline
\textbf{Parameter to be constrained} & \textbf{Source} & \textbf{Description} & \textbf{Constraint}\\
\hline
{\footnotesize Density parameter of the field at equality} & {\footnotesize EDE literature} \cite{notexcluded} & {\footnotesize Upper limit such that structure formation is not impeded and lower limit such that EDE actually has an effect} & $0.015\leq\mathit{\Omega}_{\phi}^{\textnormal{eq}}< 0.107$ \\
\hline
{\footnotesize Density parameter of the field at last scattering} & {\footnotesize EDE literature} \cite{howearly} & {\footnotesize Upper bound such that EDE cannot currently be detected in the CMB} & $\mathit{\Omega}_{\phi}^{\textnormal{ls}}<0.015$\\
\hline
{\footnotesize Density parameters of the field at last scattering and equality} & {\footnotesize Theoretical} & {\footnotesize Consistency check} & $\mathit{\Omega}_{\phi}^{\textnormal{eq}}>\mathit{\Omega}_{\phi}^{\textnormal{ls}}$\\
\hline
{\footnotesize Density parameter of the field today} & {\footnotesize Planck 2018 \cite{Planck2018}} & {\footnotesize Observational} & $0.6833\leq\mathit{\Omega}_{\phi}^{0}\leq0.6945$ \\
\hline
{\footnotesize Barotropic parameter of the field today} & {\footnotesize Planck 2018 \cite{Planck2018}} & {\footnotesize Observational} & $-1\leq w^{0}_{\phi} \leq -0.95$ \\
\hline
{\footnotesize Running of the barotropic parameter today} & {\footnotesize Planck 2018 \cite{Planck2018}} & {\footnotesize Observational} & $-0.55 \leq w^{a}_{\phi}
\leq 0.03$ \\
\hline
{\footnotesize Hubble constant in units of $\text{km/s/Mpc}$} & {\footnotesize SH0ES 2021 \cite{sh0es2021}} & {\footnotesize Observational} & $72.00\!\leq\! H_0 \!\leq\! 74.08$\\
\hline
{\footnotesize Total Field Excursion} & {\footnotesize Theoretical} & {\footnotesize 
Sub-Planckian field excursion to minimise fifth-force problems and avoid excessive radiative corrections to the potential} & $\phi_{0} - \phi_{\textnormal{eq}}<\m$\\
\hline
 
\end{tabular}
\caption{Table describing and justifying constraints used to identify the viable parameter space. In the above, \mbox{$w^{a}_{\phi} = -\left.\frac{{\rm d}w_\phi}{{\rm d}a}\right|_0$}, \textit{cf.} Eq.~\eqref{CPL}.}
\label{tab:constraints}

\end{table}

\section{Results and analysis}
\label{subsec:resultanalysis}

\subsection{Parameter Space}
\label{subsec:paramspace}

\begin{figure}[H]
    \centering
    \includegraphics[width=0.7\textwidth]{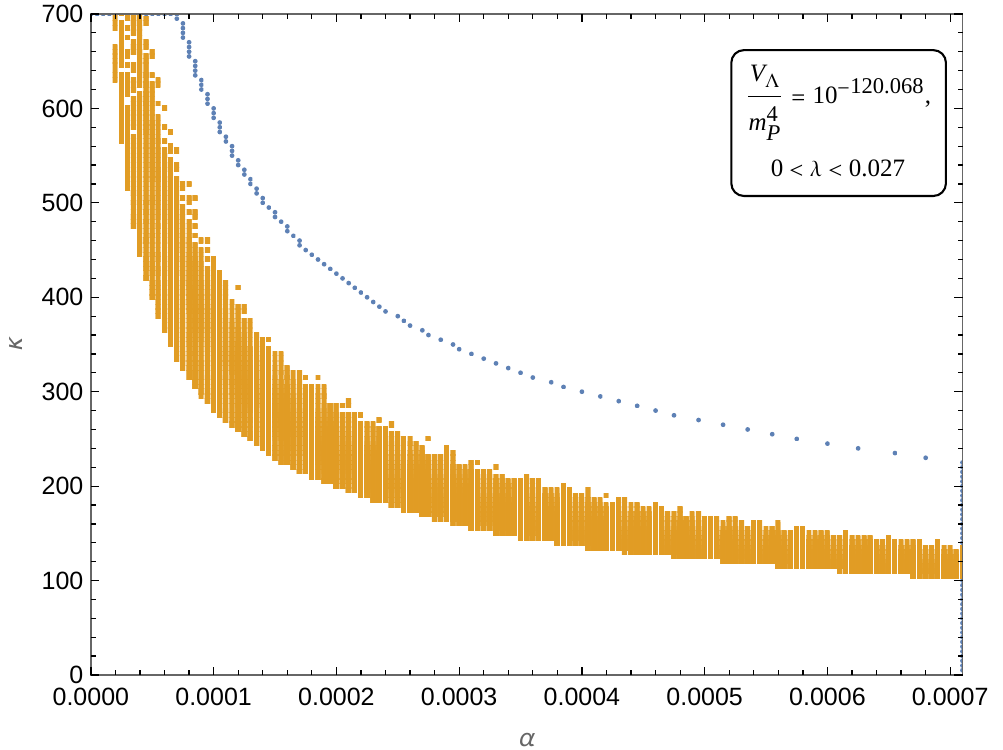}
    \caption{Parameter space slice in the $\kappa-\alpha$ plane with $0<\lambda<0.027$ and $V_{\Lambda}=10^{-120.068}\m^4$. The blue dotted line is the boundary of the region that produces non-inflationary results (see below), while the orange region is constituted by the successful points, \textit{i.e.}, those for which the constraints detailed in Table \ref{tab:constraints} are satisfied.
    Note that the region bounded in blue is not equal to the range of the scan, which is $0\leq\kappa\leq 700$ \text{and}  $0 \leq \alpha \leq 0.00071$. This is because points with potential larger than a certain starting value result in the field beginning the simulation dominant, which means that the Universe goes into inflation which cannot terminate and will never lead to successful EDE. These points are very close to the viable parameter space for these two parameters and therefore must be thrown away.}
    \label{fig:kappaalpha}
\end{figure}

\begin{figure}[h]
     \centering
     \begin{subfigure}[b]{0.495\textwidth}
         \centering
         \includegraphics[width=\textwidth]{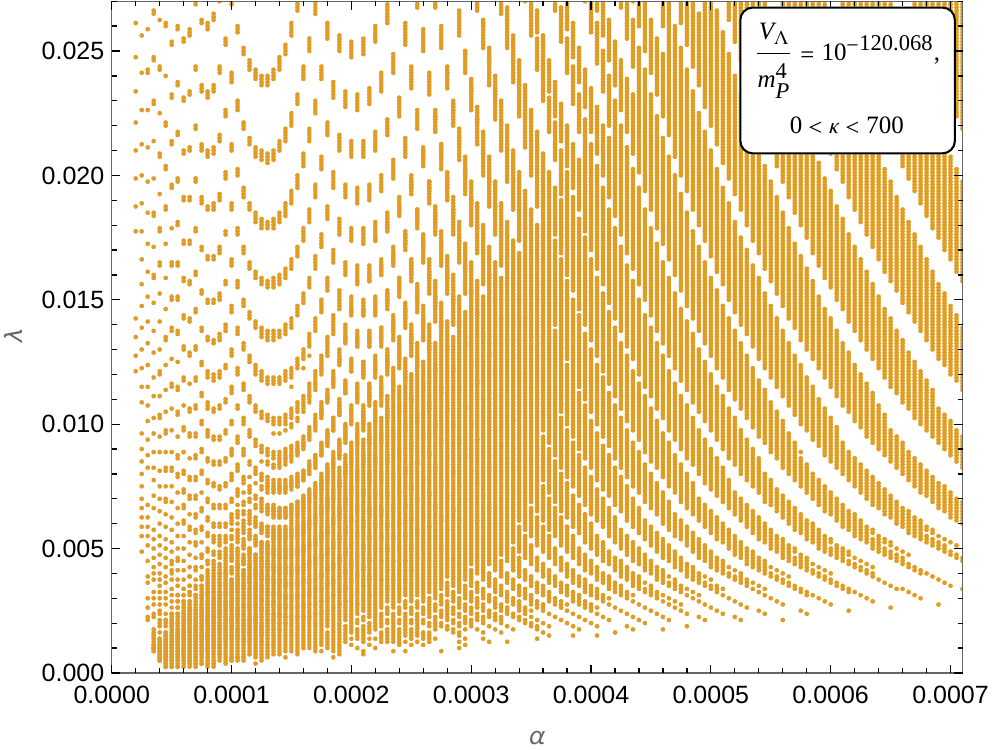}
     \end{subfigure}
     \begin{subfigure}[b]{0.495\textwidth}
         \centering
         \includegraphics[width=\textwidth]{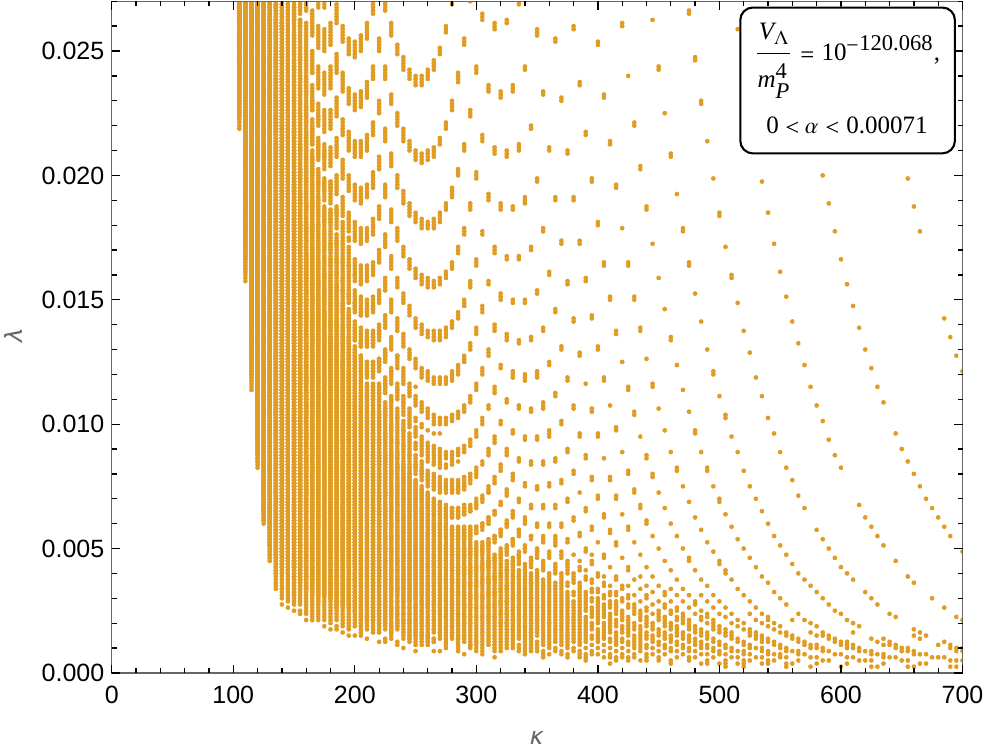}
     \end{subfigure}
     \caption{Parameter space slice in the $\lambda-\alpha$ plane with $0<\kappa<700$ (left) and in the $\lambda-\kappa$ plane with $0<\alpha<0.00071$ (right), both with $V_{\Lambda}=10^{-120.068}\m^4$. The orange region is constituted by the successful points, \textit{i.e.}, those for which the constraints detailed in Table \ref{tab:constraints} are satisfied.}
     \label{fig:lambdaalpha}
\end{figure}

We perform a scan of the parameter space of the theory, at the background level, imposing the conditions in Table \ref{tab:constraints}. We report our findings in Fig.~\ref{fig:kappaalpha}, Fig.~\ref{fig:lambdaalpha}. We find that our model is succesful for $\kappa \sim 10^{2}$ and $\lambda \sim 10^{-3} - 10^{-2}$, which are rather reasonable values. In particular, the value of $\kappa$ suggests that the mass-scale which suppresses the non-canonical field $\varphi$ in the original potential in Eq.~\eqref{Vvarphi} is near the scale of grand unification $\sim 10^{-2}\,\m$. Regarding the curvature of field space we find $\alpha \sim 10^{-4}$,  which again is not unreasonable.

The viable parameter space suggests that $\kappa\lambda>\sqrt{3}$, which contradicts our assumption in Eq.~\eqref{z}. This implies that, unlike the analytics in Sec.~\ref{subsubsec:expectedbehaviour}, the field does not adopt the subdominant exponential scaling attractor but the slow-roll exponential attractor, which leads to domination \cite{Copeland:1997et,kostasbook}. As the field thaws and starts following this attractor, the approximation in Eq.~\eqref{eqn:lowphiapprox} breaks down
as the field experiences the full exp(exp) potential, which is steeper than the
exponential (see Fig.~\ref{fig:canonicalapproximations}). Consequently, instead of becoming dominant the field free-falls. This contradiction with our discussion in 
Sec.~\ref{subsubsec:expectedbehaviour} is not very important. The existence of the scaling attractor provided an easy analytic estimate for the moment when the field unfreezes. It turns out that, because the scaling attractor has been substituted by the slow-roll attractor, 
the field unfreezes because its potential energy density becomes comparable to the total energy density, going straight into free-fall. It is much harder to analytically estimate when exactly this takes place, but the eventual result (free-fall) is the same.

We obtain that the matter-radiation equality redshift is \mbox{$z_{\rm eq}\simeq 4000$}, larger than the Planck value $z_{\rm eq}=3387\pm21$ \cite{Planck2018}. It should be however noted that, in our simplified background analysis, we use the Planck matter density parameter $\mathit{\Omega}_{\rm m.0}=0.3111\pm0.0056$ with the SH0ES value for the Hubble constant $H_0=73.04\pm1.04\,\text{km/s/Mpc}$, which is bound to give a value for $\omega_{\rm m}=\mathit{\Omega}_{\rm m,0}h^2$ incompatible with Planck. A simple back-of-the-envelope calculation shows that there is a factor of 
$\left(\frac{h_{\textnormal{SH}0\textnormal{ES}}}{h_{\textnormal{Planck}}}\right)^{2} = \left(\frac{0.73}{0.67}\right)^{2}=1.187$ difference, which leads to a new $z_{\rm eq}^{\textnormal{updated}}+1 = (1.18)^{1/3}(z_{\rm eq}+1)$, \textit{i.e.}, resulting in $z_{\rm eq}^{\rm updated}\simeq 3500$. This pushes $z_{\rm eq}$ to higher values, closer to our findings. We emphasize, however, that a full fit to the CMB data is required in order to obtain the actual value for $z_{\rm eq}$ derived from our model. In contrast, the redshift of last scattering is where we would expect it at \mbox{$z_{\rm ls}\simeq 1087$}. Theoretical constraints suggest $z_{\rm ls} \simeq 1090$ \cite{lastscatteringtheory}, and the observations of the Planck satellite suggest \mbox{$z_{\rm ls} = 1089.80\pm 0.21$}~\cite{Planck2018}. We note here that the best-fit values for the cosmological parameters from $\Lambda$CDM are expected to somewhat change when incorporating EDE. In this way, the constraints in Table \ref{tab:constraints} should be considered as approximate only.

\subsection{Field Behaviour}
\label{subsec:fieldbehaviour}
\begin{figure}[h]
     \centering
     \begin{subfigure}[b]{0.495\textwidth}
         \centering
         \includegraphics[width=\textwidth]{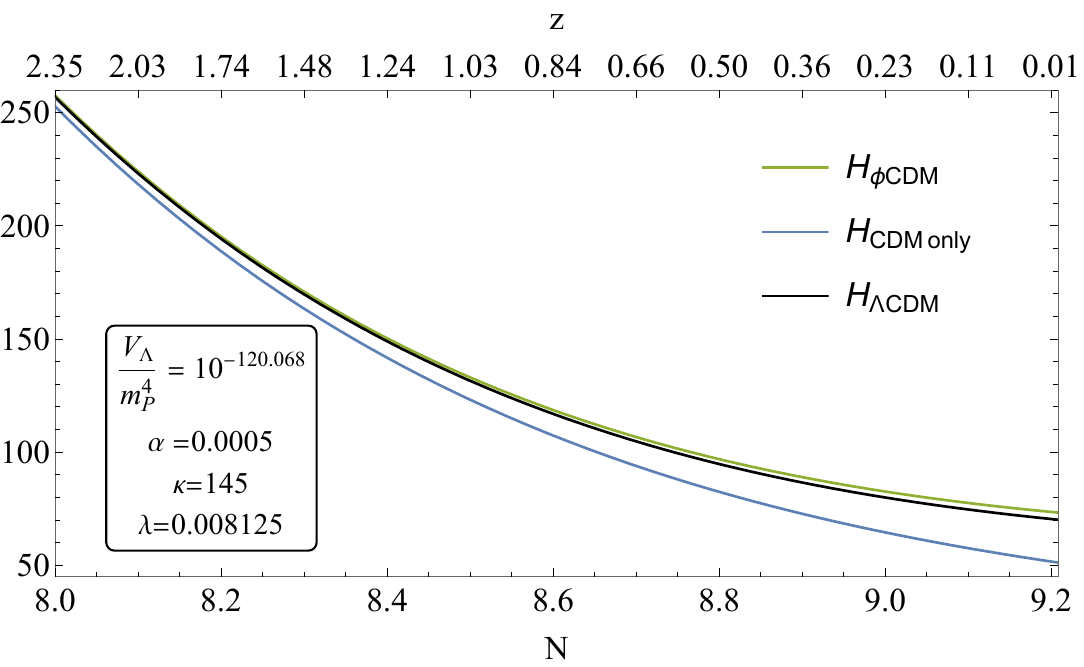}
     \end{subfigure}
     \begin{subfigure}[b]{0.495\textwidth}
         \centering
         \includegraphics[width=\textwidth]{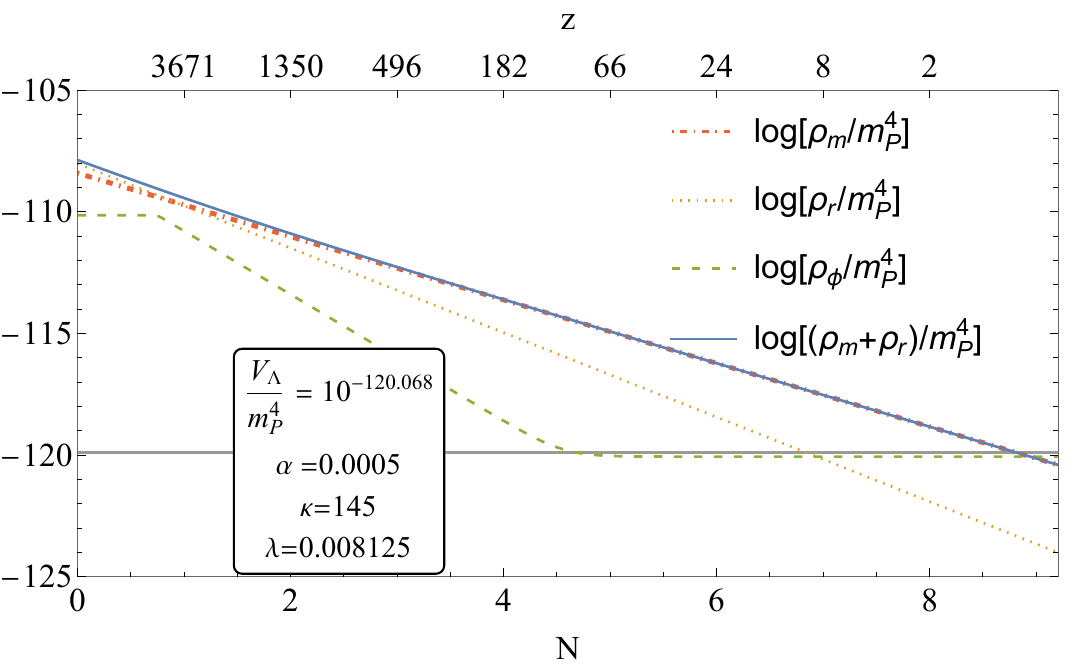}
     \end{subfigure}
     \caption{Left: The Hubble parameter (in units of $ \textnormal{km} \ \textnormal{s}^{-1}\textnormal{Mpc}^{-1}$) of a universe with  an EDE/quintessence field (green), a $\Lambda$CDM universe (black), and one with only matter and radiation (blue), as a function of redshift (top) and e-folds (bottom) elapsed since the beginning of the simulation. The presence of the field leads to a higher value of $H_0$ than in the \mbox{$\rm\Lambda$}CDM scenario. Right: The logarithmic densities of matter (dot-dashed red), radiation (dotted orange), the sum of both (solid blue) and the scalar field (dashed green), as a function of redshift (top) and e-folds (bottom) elapsed since the beginning of the simulation, for $\alpha=0.0005, \ \kappa=145, \ \lambda=0.008125$, and $V_{\Lambda}=10^{-120.068}\m^4$. The horizontal solid line represents the SH0ES energy density of the Universe at present. The EDE scalar field becomes momentarily subdominant near equality, then redshifting away faster than radiation to become negligible at decoupling.}
     \label{fig:hubbleanddensities}
\end{figure}

\begin{figure}[h]
     \centering
     \begin{subfigure}[b]{0.495\textwidth}
         \centering
         \includegraphics[width=\textwidth]{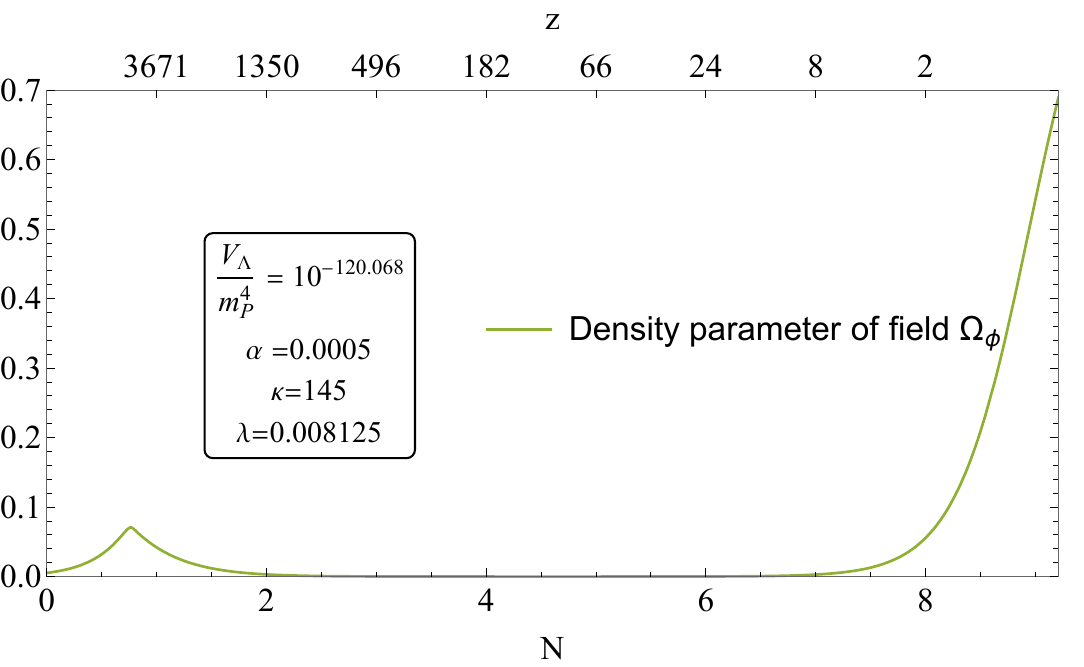}
     \end{subfigure}
     \begin{subfigure}[b]{0.495\textwidth}
         \centering
         \includegraphics[width=\textwidth]{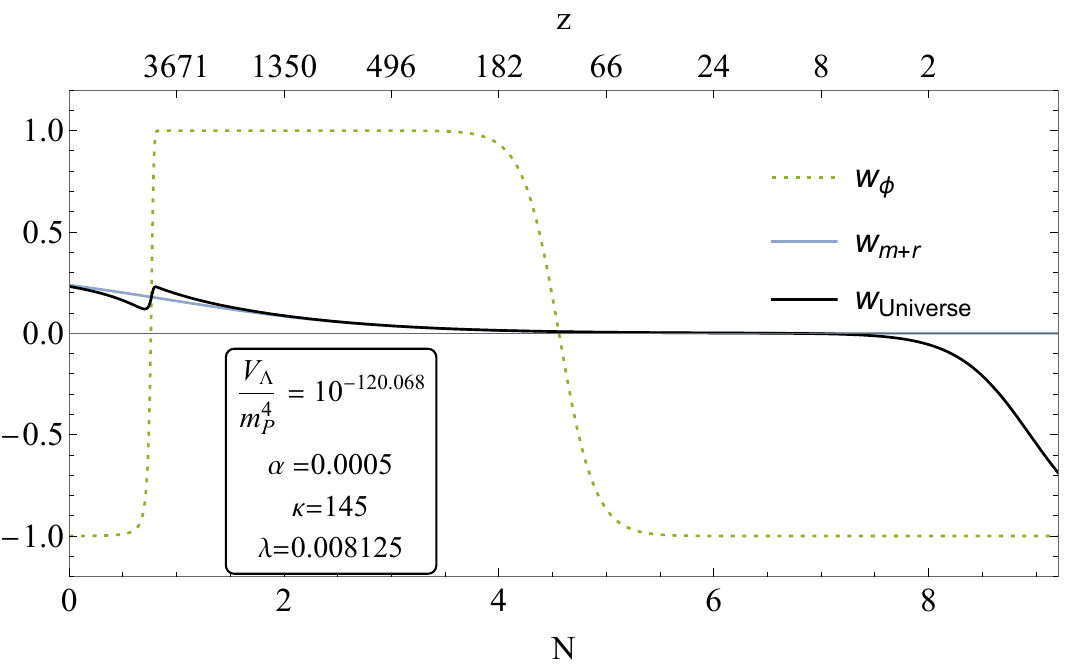}
     \end{subfigure}
     \caption{Left: The density parameter of the scalar field, for $\alpha=0.0005, \ \kappa=145, \ \lambda=0.008125$, and $V_{\Lambda}=10^{-120.068}\m^4$, as a function of redshift (top) and e-folds (bottom) elapsed since the beginning of the simulation. The density parameter experiences a bump with $f_{\textnormal{EDE}}=\mbox{$\rm \mathit{\Omega}$}_{\phi}(z_{\textnormal{eq}})\lesssim 0.1$, before the EDE redshifting away and refreezing to become dark energy today. Right: Barotropic parameter of the scalar field (dotted green), of the background perfect fluid (solid blue) and of the sum of both components (solid black), for $\alpha=0.0005, \ \kappa=145, \ \lambda=0.008125$, and $V_{\Lambda}=10^{-120.068}\m^4$. It is apparent that the scalar field becomes immediately kinetically dominated ($w_{\phi}=1$) after thawing, remaining in freefall until it refreezes again.}
    \label{fig:densityparameterandbarotropicparameter}
\end{figure}

The field behaves as expected, with the mild modification of the attractor solution at unfreezing
(slow-roll instead of scaling), which leads to free-fall. 
The evolution is depicted in Fig.~\ref{fig:hubbleanddensities}, Fig.
\ref{fig:densityparameterandbarotropicparameter}
for the example point at $\alpha=0.0005, \ \kappa=145, \ \lambda=0.008125$, and $V_{\Lambda}$ tuned to the $\textnormal{SH}0\textnormal{ES}$ cosmological constant \cite{sh0es2021}. 
The observables obtained in this case
(\textit{i.e.} the values of $H_0$, $w_0$ and $w_a$) are shown in Table~\ref{tab:fieldpoint}. The behaviour of the Hubble parameter is a function of redshift as can be seen in the left panel of Fig.~\ref{fig:hubbleanddensities}.

\begin{table}[h]

\centering
\begin{tabular}{|p{5.8cm}|p{3cm}|}
\hline
Constraint & \mbox{Example Value}\\ 
\hline
$0.015\leq\mathit{\Omega}_{\phi}^{\textnormal{eq}}< 0.107$ & 0.05178\\
& \\
$\mathit{\Omega}_{\phi}^{\textnormal{ls}}<0.015$ & 0.001722\\
& \\
$\mathit{\Omega}_{\phi}^{\textnormal{eq}}>\mathit{\Omega}_{\phi}^{\textnormal{ls}}$ & YES\\
& \\
$0.6833\leq\mathit{\Omega}_{\phi}^{0}\leq0.6945$ & 0.6889\\
& \\
$-1\leq w^{0}_{\phi} \leq -0.95$ & -1.000\\
& \\
$-0.55 \leq w_{\phi}^{a} 
\equiv -\left.\frac{{\rm d}w_{\phi}}{{\rm d}a}\right|_0 
\leq 0.03$ & $-4.850 \times 10^{-11}$\\
& \\
$72.00\leq\frac{\mbox{\normalsize $H_{0}$}}{\mbox{$\rm km\; s^{-1}\,Mpc^{-1}$}}\leq 74.08$ & {\bf 
73.27}\\
& \\
$\kappa\lambda$ & 1.178\\ 
& \\
$(\phi_{0} - \phi_{\textnormal{eq}})/\m<1$
 & 0.4274\\ 
\hline
\end{tabular}
\caption{Table giving the constraints and their corresponding values for an example point, $\alpha=0.0005, \ \kappa=145, \ \lambda=0.008125$, and $V_{\Lambda}$ tuned to the $\textnormal{SH}0\textnormal{ES}$ cosmological constant, in the viable parameter space. The Hubble constant obtained in this example is \mbox{$H_0=73.27\,$km/s\ Mpc.}
}
\label{tab:fieldpoint}
\end{table}

As shown in Table \ref{tab:constraints}, the maximum allowed value of the EDE density parameter at equality is just over 0.1. However, it is possible that this is too lenient a constraint because unlike the models for which this constraint was developed, our model has a true free-fall period, which means it redshifts away \textit{exactly} as $a^{-6}$ rather than below this rate as in oscillatory behaviour (see the right panels of Fig.~\ref{fig:hubbleanddensities}, Fig. \ref{fig:densityparameterandbarotropicparameter})\footnote{A more accurate constraint of $\sim0.086$ for non-oscillatory models is provided in Ref.~\cite{Lin_2019}, which does not significantly narrow our allowed parameter space.}. Note that for oscillating EDE in a potential $V\propto \phi^{2n}$, as the original EDE \cite{edecanresolve}, there is a limit $n<3$ ($n<5$) for matter (radiation) domination. This is because for $n>3$ ($n>5$) there exists an scaling attractor $\phi\propto t^{1/(1-n)}$, which means that oscillations are impeded \cite{PhysRevD.37.3406,Liddle:1998xm}. Recently, a similar result was found in Ref. \cite{Smith:2019ihp}, where it is shown that the data favours $2\lesssim n \lesssim 3.4$ at the $68\%$ C.L. Since EDE typically unfreezes around matter-radiation equality, this implies that the density of oscillating EDE cannot decrease faster than $\rho_{\phi}\propto a^{-9/2}$, \textit{i.e.}, not as fast as true free-fall, where $\rho_\phi\propto a^{-6}$ as we obtain.

At present, the exponential contribution to the potential density in Eq.~\eqref{eqn:highphiapprox} is largely subdominant to $V_\Lambda$, 
so the contribution of the scalar field to the total density budget is almost constant, as in $\rm\Lambda$CDM. Its barotropic parameter is, therefore, \mbox{$w_\phi\approx -1$}
(see the right panel of Fig.~\ref{fig:densityparameterandbarotropicparameter}). Technically, it is not exactly -1 but its running is negligible, with the viable parameter space for $w_a$ fitting easily within the constraint in Eq.~\eqref{CPL} by some ten orders of magnitude (see Table~\ref{tab:fieldpoint}).

\section{Initial Conditions}
\label{sec:discussion}

Our model accounts for both EDE and late-time dark energy in a non-oscillatory manner (in contrast to Ref.~\cite{alphaattractorsede1}).
The field is frozen at early times, thawing just before matter-radiation equality when its density grows to nearly 0.1 of the total value (see left panel of Fig.~\ref{fig:densityparameterandbarotropicparameter}), as set by constraints in Ref.~\cite{notexcluded}. A steep $\exp(\exp)$ potential then forces the field into free-fall, causing its energy density to dilute away as $\rho_{\phi} \propto a^{-6}$. After this, the field hits the asymptote of the exponential decay and refreezes, becoming dominant at present (see the right panel of Fig.~\ref{fig:hubbleanddensities}).

Thus, we achieve DE-like behaviour at the present day by ensuring that the field refreezes after its period of free-fall, therefore remaining at a constant energy density equal to the value of the potential density at that point. Although this constant potential density is initially negligible, the expansion of the Universe causes the density of matter to decrease. Because the field refreezes at a potential density that is comparable to the density of matter at present, the field starts to become dominant at the present day. Once it begins to dominate the Universe, the field thaws again, but the density of the Universe is dominated by a constant contribution $V_\Lambda$, as with $\rm\Lambda$CDM.

The obvious question is why our scalar field finds itself frozen at the origin in the first place. One compelling explanation is the following.

We assume that the origin is an enhanced symmetry point (ESP) such that, at very early times, an interaction of $\varphi$ with some other scalar field $\chi$ traps the rolling of $\varphi$ at zero. The idea follows the scenario explored in Ref.~\cite{Kofman:2004yc}.
In this scenario, the scalar potential includes the interaction
\begin{equation}
    \Delta V=\frac12g^2\varphi^2\chi^2\,,
    \label{DV}
\end{equation}
where the coupling $g<1$ parametrises the strength of the interaction. Note that here $\varphi$ is the non-canonical scalar field, appearing in the Lagrangian in Eq. \eqref{L0}, related to its canonical version $\phi$ via Eq. \eqref{phivarphi}. It is also featured in our potential, when it is first introduced in Eq.~\eqref{Vvarphi}.

We assume that initially $\varphi$ is rolling down its steep potential\footnote{Far away from the origin, the scalar potential $V(\varphi)$ does not have to be of the form in Eq.~\eqref{Vvarphi}. In fact, it is conceivable that $\varphi$ might play the role of the inflaton field too (see \ref{appendix:A}).}. Then, the interaction in Eq.~\eqref{DV} provides a modulated effective mass-squared \mbox{$m_{\rm eff}^2=g^2\varphi^2$}
to the scalar field $\chi$. When $\varphi$ crosses the origin, this effective mass becomes momentarily zero. If the variation of the $\varphi$ field (\textit{i.e.} the speed $|\dot\varphi|$ in field space) is large enough, then there is a window around the origin when \mbox{$|\dot{m}_{\rm eff}|\gg m_{\rm eff}^2$} (because, \mbox{$|\dot\varphi|\gg\varphi^2\simeq 0$}). This violates adiabaticity and leads to copious production of $\chi$-particles \cite{Kofman:2004yc}\footnote{Near the origin, when \mbox{$\varphi\simeq 0$}, the $\varphi$-field is approximately canonically normalised, as suggested by Eq.~\eqref{phivarphi}, so the considerations of Ref.~\cite{Kofman:2004yc} are readily applicable.}. 

As the field moves past the ESP, the produced $\chi$ particles become heavy, which takes more energy from the $\varphi$ field, producing an effective potential incline in the direction the $\varphi$ field is moving. Indeed, the particle production generates an additional linear potential \mbox{$\sim g|\varphi|n_\chi$} \cite{Kofman:2004yc}, where $n_\chi$ is the number density of the produced $\chi$-particles. This number density is constant because the duration of the effect is much smaller than a Hubble time, so that we can ignore dilution from the Universe expansion. The rolling $\varphi$ field climbs up the linear potential until its kinetic energy density is depleted. Then the field momentarily stops and afterwards reverses its motion (variation) back to the origin. When crossing the origin again, there is another bout of $\chi$-particle production, which increases $n_\chi$ and makes the linear potential steeper to climb. This time, $\varphi$ variation halts at a value closer to the origin. Then, the field reverses its motion and rushes through the origin again. Another outburst of $\chi$-particle production steepens the linear potential further. The process continues until the $\varphi$-field is trapped at the origin \cite{kostasbook,Kofman:2004yc}.

The trapping of a rolling scalar field at an ESP can take place only if the $\chi$-particles do not decay before trapping occurs. 

If they did, the $n_\chi$ would decrease and the potential $g|\varphi|n_\chi$ would not be able to halt the motion (variation) of the $\varphi$-field. The end result of this process is that all the kinetic energy density of the rolling $\varphi$ has been given to the $\chi$-particles. Now, since $\varphi$ is trapped at the origin, the effective mass of the $\chi$-particles is zero, which means that they are relativistic matter, with density scaling as 
\mbox{$\rho_\chi\propto a^{-4}$}. As far as $\varphi$ is concerned, it is trapped at the origin and its density is only
\mbox{$\rho_\varphi=V(\varphi=0)=e^{-\lambda} V_X=\,$constant} (\textit{cf.}~Eq.~\eqref{Vvarphi}).

\begin{figure}[H]
    \centering
    \includegraphics[width=0.6\textwidth]{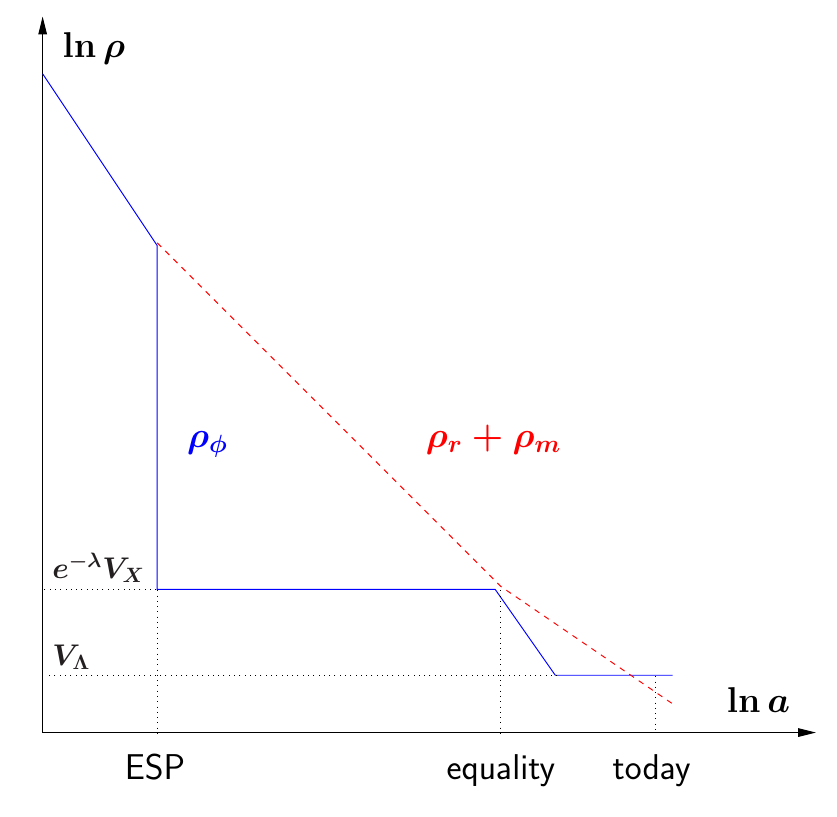}
    \caption{Schematic log-log plot depicting the evolution of the density of the scalar field $\rho_\phi$ (solid blue line) and the density of radiation and matter $\rho_r+\rho_m$ (dashed red line) in the case when the decay of the kinetic energy density of the trapped scalar field generates the thermal bath of the hot Big Bang (as in Ref.~\cite{Dimopoulos:2019ogl}). Originally the $\phi$-field is rushing towards the minimum of the potential, dominated by its kinetic density, so that \mbox{$\rho_\phi\propto a^{-6}$} (free-fall). When it crosses the enhanced symmetry point (ESP) its interaction to the $\chi$-field (\textit{cf.} Eq.~\eqref{DV}) traps the rolling $\phi$-field at the ESP while all its kinetic energy is given to $\chi$-particles, which soon decay into the radiation and matter of the hot Big Bang (the decay is assumed to be quick, just after trapping). Afterwards, the $\phi$-field stays frozen, with energy density \mbox{$V(\phi=0)=e^{-\lambda}V_X$} (\textit{cf.}~Eq.~\eqref{Vvarphi}) until much later, when its potential density is comparable to the background. Then it unfreezes before dominating, acting as early dark energy at the time near matter-radiation equality, and subsequently free-falls to its value $\phi_0$, with potential density approximately \mbox{$V_\Lambda=\,$constant}. The field stays there until the present when it dominates the Universe and becomes late dark energy.}
    \label{fig:ede}
\end{figure}

After some time, it may be assumed that the $\chi$-particles do eventually decay into the standard model particles, which comprise the thermal bath of the hot Big Bang. The confining potential, which is proportional to $n_\chi$, disappears but, we expect the $\varphi$-field to remain frozen at the origin because the scalar potential $V(\varphi)$ in Eq.~\eqref{Vvarphi} is flat enough there. As we have discussed, the $\varphi$-field unfreezes again in matter-radiation equality. The above scenario is depicted in Fig.~\ref{fig:ede}

For simplicity, we have considered that, apart from the obvious violation of adiabacity at the ESP, the $\chi$ direction is otherwise approximately flat and the $\chi$-field has a negligible bare mass compared to the $\varphi$ field. It would be more realistic to consider a non-zero bare mass for the $\chi$-particles, which when they become non-relativistic (much later than the trapping of $\varphi$) can safely decay to the thermal bath of the hot Big Bang, reheating thereby the Universe, \textit{e.g.} in a manner not dissimilar to Ref.~\cite{Dimopoulos:2019ogl}.

The above scenario is one possible explanation of the initial condition considered and not directly relevant to the scope of this work - we simply assume that the field begins frozen at the origin. Other possibilities to explain our initial condition exist, for example considering a thermal correction of the form \mbox{$\delta V\propto T^2 \varphi^2$}, which would make the origin an effective minimum of the potential at high temperatures and drive the $\varphi$-field there.

\section{Conclusions}
\label{sec:conclusions}

In conclusion, we have proposed a toy model that unifies EDE and DE via a scalar field in the context of $\alpha$-attractors. We have studied the background dynamics in detail, finding that the value of the Hubble parameter, coming from early-time data, can be raised while simultaneously explaining the current accelerated expansion, with no more fine tuning than $\rm\Lambda$CDM.

Our work differs from Ref. \cite{alphaattractorsede1}, in  that the field is not oscillating; instead after equality, it free-falls with energy density decreasing as $\rho\propto a^{-6}$, faster than most EDE proposals and the fastest possible\footnote{Causality implies that the barotropic parameter $w$ of a perfect fluid cannot be larger than unity because the speed of sound of the fluid \mbox{$c_s^2=w$} cannot be superluminal. This implies $w\leq 1$ and so, the density of an independent perfect fluid \mbox{$\rho\propto a^{-3(1+w)}$} cannot decrease faster than $a^{-6}$. However, a homogeneous scalar field can be represented as a perfect fluid with \mbox{$w=\frac{\rho_{\rm kin}-V}{\rho_{\rm kin}+V}$}, where $\rho_{\rm kin}$ is the kinetic energy density of the scalar field and $V$ the potential. It seems that $w>1$ could indeed happen when the field transverses an AdS minimum of $V$, such that $V<0$. As a result, the density of such scalar field could decrease faster than $a^{-6}$. The scenario of such EDE has been considered in Refs.~\cite{Ye:2020oix,Ye:2020btb}.}. Although, from our background analysis, we find a larger value of $z_{\rm eq}$ than found by Planck, it should be realised that Planck assumes a $\rm\Lambda$CDM scenario to derive this quantity and hence it may not be fully applicable to other models, particularly one with a significant scalar field contribution at that time as in our case. Of course, a full fit to the CMB data is needed in order to obtain the actual $z_{\rm eq}$ derived from our model.

In our proposed scenario, 
the scalar field lies originally frozen at the origin, until it thaws near the time of equal matter-radiation densities, when it becomes EDE. Afterwards it free-falls until it refreezes at a lower potential energy density value, which provides the vacuum density of $\rm\Lambda$CDM. We showed that the total excursion of the field in configuration space is sub-Planckian, which implies that our potential is stable under radiative corrections.

One explanation of our initial conditions is that the origin is an ESP. Our scalar field is originally kinetically dominated until it is trapped at the ESP when crossing it\footnote{A thermal correction to the scalar potential can have a similar effect.}. As we discuss in \ref{appendix:A}, the scalar field could even be the inflaton, which after inflation rolls down its runaway potential until it becomes trapped at the ESP.

 Our potential in Eq.~\eqref{Vvarphi} really serves to demonstrate that a model unifying EDE with $\rm\Lambda$CDM can be achieved with a suitably steep runaway potential. With the parameters of our model assuming rather natural values, thereby not introducing fine-tuning additional to that of $\rm\Lambda$CDM, we show that this is indeed possible with a simple design. 

The challenge lies in constructing
a concrete theoretical framework for such a potential. Furthermore, although the background analysis is promising, a full fit to the CMB data is lacking. We plan on running a Markov Chain Monte Carlo (MCMC) doing this in a future work. This is of paramount importance since it would show what values (if any) from our \textit{a priori} viable parameter space lead to a best fit to the data.

\section*{Acknowledgements}
LB is supported by STFC. KD is supported (in part) by the Lancaster-Manchester-Sheffield Consortium for Fundamental Physics under STFC grant: ST/T001038/1. SSL is supported by the FST of Lancaster University. For the purpose of open access, the authors have applied a Creative Commons Attribution (CC BY) licence to any Author Accepted Manuscript version arising.

\appendix

\section{Quintessential Inflation} \label{appendix:A}

Is it possible that our scalar field can not only be early and late dark energy, but also be the inflaton field, responsible for accelerated expansion in the early Universe?

The $\alpha$-attractors construction leads to two flat regions in the scalar potential of the canonical field, as the kinetic poles of the non-canonical field are displaced to infinity. This idea has been employed in the construction of quintessential inflation models in Refs.~\cite{Akrami:2017cir,Dimopoulos:2017tud,Dimopoulos:2017zvq}, where the low-energy plateau was the quintessential tail, responsible for quintessence and the high-energy plateau was responsible for inflation.

However, if we inspect the potential in Eq.~\eqref{Vvarphi} at the poles \mbox{$\varphi=\pm\sqrt{6\alpha}\,\m$}, we find that the potential for the positive pole is \mbox{$V(\varphi_+)=V_\Lambda$} as expected, while for the negative pole we have \mbox{$V(\varphi_-)=V_\Lambda\exp[2\lambda\sinh(\kappa\sqrt{6\alpha})]$}. For the values of the parameters obtained (\mbox{$\kappa\sim 10^2$}, \mbox{$\lambda\sim 10^{-3}$} and \mbox{$\alpha\sim 10^{-4}$}) it is easy to check that $V(\varphi_-)$
is unsuitable for the inflationary plateau. Thus, our model needs to be modified to lead to quintessential inflation.

The first modification is a shift in field space such that our new field is
\begin{equation}
    \tilde\varphi=\varphi+\Phi\,,
    \label{tilde}
\end{equation}
where $\Phi$ is a constant. The $\alpha$-attractors construction applies now on the new field $\tilde\varphi$ for which the Lagrangian density is given by the expression in Eq.~\eqref{L0} with the substitution \mbox{$\varphi\rightarrow\tilde\varphi$}. The poles of our new field lie at \mbox{$\tilde\varphi_\pm=\pm\sqrt{6\tilde\alpha}\,\m$}, where $\tilde\alpha$ is the new $\alpha$-attractors parameter.

We want all our results to remain unaffected, which means that, for the positive pole, Eq.~\eqref{tilde} suggests
\begin{equation}
\begin{split}
    \varphi_+ & =\sqrt{6\alpha}\,\m=\tilde\varphi_+-\Phi=\sqrt{6\tilde\alpha}\,\m-\Phi\;\\
    \Rightarrow\;&
    \tilde\alpha=\frac16\left(\frac{\Phi}{\m}+\sqrt{6\alpha}\right)^2\,.
    \label{alphatilde}
\end{split}
\end{equation}

The above, however, is not enough. It turns out we need to modify the scalar potential as well. This modification must be such that near the positive pole the scalar potential reduces to the one in Eq.~\eqref{Vvarphi}. A simple proposal is
\begin{equation}
    V(\tilde\varphi)=V_X\exp\{-2\lambda\sinh[\kappa(\tilde\varphi-\Phi)/\m]\}\,,
    \label{Vtildevarphi}
\end{equation}
which indeed reduces to Eq.~\eqref{Vvarphi} when \mbox{$\kappa(\tilde\varphi-\Phi)=\kappa\varphi>\m$}.
Note that \mbox{$\kappa\sqrt{6\alpha}>1$} is implied from the requirement that near the positive pole
we have \mbox{$\kappa\sqrt{6\alpha}\,\m=\kappa\varphi_+>\m$}.

The ESP discussed in Sec.~\ref{sec:discussion} is now located at \mbox{$\tilde\varphi=\Phi$}, such that Eq.~\eqref{DV} is now \mbox{$\Delta V=\frac12g^2(\tilde\varphi-\Phi)^2\chi^2$}.\footnote{Near the ESP the potential does not approximate Eq.~\eqref{Vvarphi}. However, we assume that, after unfreezing, the field rolls away fast from the ESP, such that soon the exp(exp) form of the potential becomes valid and the evolution is the one discussed in the main text of our paper.}

We are interested in investigating the inflationary plateau. This is generated for the canonical field near the negative pole \mbox{$\tilde\varphi_-=-\sqrt{6\tilde\alpha}\,\m$}, where the scalar potential of the canonical field ``flattens out" \cite{lindealphaattractors}. 

Assuming that \mbox{$\Phi>\sqrt{6\alpha}\,\m$}, we have that \mbox{$\tilde\varphi_--\Phi=-2\Phi-\sqrt{6\alpha}\,\m\simeq -2\Phi$}, where we used Eq.~\eqref{alphatilde}. Hence, for the potential energy density of the inflationary plateau we obtain

\begin{flalign}
    &V_{\rm inf}=V(\tilde\varphi_{-}) &\simeq &
    V_X\exp[-2\lambda\sinh(-2\kappa\Phi/\m)] &\nonumber\\
    && \simeq & \exp(\lambda\,e^{\kappa\sqrt{6\alpha}})V_\Lambda\exp[\lambda\exp(2\kappa\Phi/\m)] &\nonumber\\
    && = & \exp[\lambda(e^{\kappa\sqrt{6\alpha}}+e^{2\kappa\Phi/\m})]V_\Lambda\simeq V_\Lambda\exp(\lambda\,e^{2\kappa\Phi/\m})\,,&
    \label{Vinf}
\end{flalign}
where we used Eq.~\eqref{Vvarphi} and 
that in \mbox{$-2\sinh (-x)\simeq e^x$}, when \mbox{$x\gg 1$}.

With $\alpha$-attractors, the inflationary predictions are \mbox{$n_s=1-2/N$} and \mbox{$r=12\tilde\alpha/N^2$} \cite{lindealphaattractors}, where $n_s$ is the spectral index of the scalar curvature perturbation and $r$ is the ratio of the spectrum of the tensor curvature perturbation to the spectrum of the scalar curvature perturbation, with $N$ being the number  of inflationary efolds remaining after the cosmological scales exit the horizon. Typically, \mbox{$N=60-65$} for quintessential inflation, which means that \mbox{$n_s=0.967-0.969$}, in excellent agreement with the observations \cite{Planck:2018jri}\footnote{It should be however noted that recent results \cite{Ye:2021nej,Jiang:2022uyg,Ye:2022efx,Jiang:2022qlj} suggest that, in the presence of EDE, the data seems to favour larger values of $n_s$, closer to unity. This would somewhat undermine the use of $\alpha$-attractors.}. For the tensor-to-scalar ratio the observations provide the bound \mbox{$r<0.036$} \cite{BICEP:2021xfz}, which suggests 
\mbox{$\tilde\alpha<0.003\,N^2=10.8-12.7$}.

The COBE constraint requires \mbox{$V_{\rm inf}\sim 10^{-10}\,\m^4$}. Using that \mbox{$V_\Lambda\sim 10^{-120}\,\m^4$}, Eq.~\eqref{Vinf} suggests that \mbox{$\kappa\Phi/\m=\frac12\ln(110\ln10/\lambda)$}. 
Hence. the conditions \mbox{$\Phi>\sqrt{6\alpha}\,\m$} and \mbox{$\kappa\sqrt{6\alpha}>1$} suggest
\begin{equation}
    1<\kappa\sqrt{6\alpha}<\kappa\Phi/\m
    = \frac12\ln(110\ln10/\lambda)\,.
    \label{COBE}
\end{equation}
Our findings in Section ~\ref{subsec:resultanalysis} are marginally in agreement with the above requirements. For example, taking \mbox{$\alpha=0.0006$} and \mbox{$\kappa=100$} we find \mbox{$\kappa\sqrt{6\alpha}=6$} and then Eq.~\eqref{COBE} suggests \mbox{$\lambda<1.556\times 10^{-3}$}. We also find \mbox{$\Phi/\m>\sqrt{6\alpha}=0.06$}, which is rather reasonable. Then, Eq.~\eqref{alphatilde} implies \mbox{$\tilde\alpha>12\alpha=7.2\times 10^{-3}$}, which comfortably satisfies the observational constraint on $r$. In fact, taking \mbox{$N\simeq 60$}, we find \mbox{$r=12\tilde\alpha/N^2>\alpha/25=
2.4\times 10^{-5}$}.

The above should be taken with a pinch of salt because the approximations employed are rather crude. However, they seem to suggest that our augmented model in Eq.~\eqref{Vtildevarphi} may lead to successful quintessential inflation while also resolving the Hubble tension, with no more fine-tuning than that of $\rm\Lambda$CDM.\footnote{Unifying inflation, EDE and late DE in $F(R)$ modified gravity has been investigated in Refs.~\cite{Nojiri:2019fft,Oikonomou:2020qah}.} A full numerical investigation is needed to confirm~this.

 \bibliographystyle{elsarticle-num} 
 \bibliography{cas-refs}

\end{document}